\documentclass[preprintnumbers,amsmath,amssymb,showpacs]{revtex4}
\usepackage{epsfig}
\usepackage{color}
\usepackage[colorlinks,urlcolor=blue,citecolor=blue]{hyperref}
\begin{document}
\setlength{\voffset}{1.0cm}
\title{Integrable Gross-Neveu models with fermion-fermion \\ and fermion-antifermion pairing}
\author{Michael Thies\footnote{michael.thies@gravity.fau.de}}
\affiliation{Institut f\"ur  Theoretische Physik, Universit\"at Erlangen-N\"urnberg, D-91058, Erlangen, Germany}
\date{\today}
\begin{abstract}
The massless Gross-Neveu and chiral Gross-Neveu models are well known examples of integrable quantum field theories
in 1+1 dimensions. We address the question whether integrability is preserved if one either replaces the four-fermion
interaction in fermion-antifermion channels by a dual interaction in fermion-fermion channels, or if one adds such a
dual interaction to an existing integrable model. The relativistic Hartree-Fock-Bogoliubov approach is adequate to deal with the large $N$ limit
of such models. In this way, we construct and solve three integrable models with Cooper pairing. We also identify a candidate
for a fourth integrable model with maximal kinematic symmetry, the ``perfect" Gross-Neveu model. This type of field theories
can serve as exactly solvable toy models for color superconductivity in quantum chromodynamics.
\end{abstract}
\pacs{11.10.Kk,11.10.-z,12.40.-y}
\maketitle
\section{Introduction}
\label{sect1}
In its original form, the Gross-Neveu (GN) model \cite{L1} is perhaps the simplest interacting, fermionic field theory. $N$ massless flavors of Dirac fermions
interact via a scalar-scalar four-fermion interaction in 1+1 dimensions,
\begin{equation}
{\cal L}_{\rm GN} = \sum_{i=1}^N \bar{\psi}^{(i)} i\partial \!\!\!/ \psi^{(i)} + \frac{g^2}{2} \left( \sum_{i=1}^N \bar{\psi}^{(i)} \psi^{(i)} \right)^2 \, .
\label{A1}
\end{equation}
The large $N$ limit is at the same time tractable and physically suggestive of higher dimensions, and we use it throughout this paper.
Since its inception 40 years ago, a number of generalizations of the GN model have been considered, adding a bare mass term or modifying
the interaction. The best known such generalization is presumably the chiral GN model, the 2d version of the even older 
Nambu--Jona-Lasinio (NJL) model \cite{L2},
\begin{equation}
{\cal L}_{\rm NJL} =  \sum_{i=1}^N \bar{\psi}^{(i)} i\partial \!\!\!/ \psi^{(i)} + \frac{g^2}{2} \left[\left( \sum_{i=1}^N \bar{\psi}^{(i)} \psi^{(i)} \right)^2 +
\left( \sum_{i=1}^N \bar{\psi}^{(i)} i \gamma_5\psi^{(i)} \right)^2\right] \, .
\label{A2}
\end{equation}
Here the discrete Z$_2$ chiral symmetry of (\ref{A1}) gets promoted to a continuous U(1) chiral symmetry. Other four-fermion interactions which
can be found in the literature interpolate between (\ref{A1}) and (\ref{A2}) by introducing two different coupling constants \cite{L3,L4,L5}, or have extra terms which give rise
to fermion-fermion pairing rather than fermion-antifermion pairing \cite{L6,L7,L8,L9,L10,L11,L12}. The motivation for the latter models comes mostly from the predicted 
phenomenon of color superconductivity in quantum chromodynamics \cite{L13,L14,L15}. Accordingly, the emphasis of these works has typically been on the phase diagram 
of the corresponding models and patterns of symmetry breaking.

One property which singles out the Lagrangians (\ref{A1},\ref{A2}) from the vast majority of their modifications is integrability \cite{L16,L17,L18}. While this does not seem to play any role
in the calculation of thermodynamic quantities, integrability permits to solve
static and even time dependent soliton problems in the massless GN and NJL models explicitly. Thus, 
scattering problems involving any number of kinks, kink-antikink baryons, compound bound states and breathers 
have been solved analytically by time-dependent Hartree-Fock methods (TDHF) recently \cite{L19,L20,L21,L22}. Nothing comparable has been 
achieved for massive GN models, or any variant with different interaction terms, for that matter, so that integrability is undoubtedly crucial here. 
Since it is quite exceptional to be able to solve both equilibrium thermodynamics and the time evolution of an interacting quantum field theory exactly, 
the question arises whether there are other (physically relevant) integrable four-fermion models. This is the main topic of the present
paper.  

This is not an easy question, therefore we shall proceed rather heuristically. Two main ingredients have proven helpful in our search for
integrability: The first one is related to symmetries, the second one to the concept of duality between fermion-fermion and fermion-antifermion pairing.

Consider symmetry issues first. If one wishes to generalize an integrable model by making it more complicated
without losing integrability, we find it plausible that it helps if the symmetry of the starting model gets enhanced in this process. 
Thus for instance, adding a mass term to the GN Lagrangian (\ref{A1}) breaks the discrete chiral symmetry and renders the model 
non-integrable. By contrast, switching on an interaction in the pseudoscalar channel of the same strength as in the scalar channel when going from
(\ref{A1}) to (\ref{A2}) enhances the chiral symmetry and maintains integrability. Notice also that the known integrable models have only one coupling constant.
It is hard to imagine that integrability can be kept if one adds more interactions with arbitrary coupling
constants. 

Turning to duality, we remind the reader of a concept introduced in Ref.~\cite{L23} and further exploited in the recent papers \cite{L11,L12}.
The duality transformation we have in mind consists in replacing fields by their complex conjugates, separately for left-handed and
right-handed fermions (this distinguishes it from charge conjugation). Thereby one can relate models with fermion-antifermion
pairing (chiral symmetry breaking) and fermion-fermion pairing (superconductivity) to each other. This concept will turn out to be important for
identifying and characterizing potentially integrable models different from (\ref{A1},\ref{A2}).

The plan of this paper reflects these introductory remarks. In Sec.~\ref{sect2}, we start with free, massless fermions and specialize the Pauli-G\"ursey
symmetry \cite{L24,L25} to 1+1 dimensions.  In Sec.~\ref{sect3}, we briefly recall the duality transformation applied
to the two integrable models (\ref{A1},\ref{A2}). This yields two distinct integrable models with Cooper pairing only. In Sec.~\ref{sect4} we take up the
concept of ``self-dual" field theories from Ref.~\cite{L23}. Here we construct the self-dual version of the GN model.  We will confirm its integrability and solve the
model completely by reducing it to the GN model. 
In doing so, we shall introduce the appropriate framework for dealing with fermion-fermion and fermion-antifermion pairing simultaneously, 
namely the Hartree-Fock-Bogoliubov (HFB) method \cite{L26}. In Sec.~\ref{sect5}, we turn to the self-dual NJL model
having maximal symmetry in a sense which will be made more precise below. We give arguments for its integrability based on
the corresponding classical fermion model. Here we have not yet been able to solve the HFB equations in any systematic fashion,
and have to leave this for the future.  Sec.~\ref{sect6} contains a brief summary and an outlook.
\section{Pauli-G\"ursey symmetry in 1+1 dimensions}
\label{sect2}
The Pauli-G\"ursey symmetry \cite{L24,L25} is the largest kinematical symmetry group of massless Dirac fermions. It combines chiral transformations
and charge conjugation with the Poincar\'{e} group. Here, we only need the special case of 1+1 dimensions. Throughout this paper, we use the following chiral 
representation of the Dirac matrices,
\begin{equation}
\gamma^0 = \sigma_1 \ , \qquad \gamma^1 =  {\rm i}\sigma_2 \ , \qquad \gamma_5 = \gamma^0
\gamma^1 = - \sigma_3 \,  .
\label{B1}
\end{equation}
The upper and lower components of the Dirac spinor then coincide with fields of definite chirality,
\begin{equation}
\psi = \left( \begin{array}{c} \psi_L \\ \psi_R \end{array} \right) = \left( \begin{array}{c} \psi_1 \\ \psi_2 \end{array} \right) \,  .
\label{B2}
\end{equation}
Furthermore, we shall use light cone coordinates in the following convention,
\begin{equation}
z=x-t, \quad \bar{z} = x+t, \quad \partial_0 = \bar{\partial}-\partial, \quad \partial_1 = \bar{\partial}+\partial \, ,
\label{B3}
\end{equation}
so that the free, massless Dirac Lagrangian becomes
\begin{equation}
{\cal L}_0 = \bar{\psi} i \partial\!\!\!/ \psi = - 2i \psi_1^* \partial \psi_1 + 2 i \psi_2^* \bar{\partial} \psi_2 \, .
\label{B4}
\end{equation}
Leaving Poincar\'{e} transformations aside, the Pauli-G\"ursey group in 1+1 dimensions can be generated by four basic (canonical)
transformations,
\begin{eqnarray}
\psi_1 & \to & {\rm e}^{{\rm i} \alpha} \psi_1 \, , \nonumber \\
\psi_2 & \to & {\rm e}^{{\rm i}\beta } \psi_2 \, , \nonumber \\
\psi_1 & \to & \psi_1^* \,  , \nonumber \\
\psi_2 & \to & \psi_2^* \,  .
\label{B5}
\end{eqnarray}
The first two lines are the continuous chiral transformations, a symmetry shared by the classical Lagrangian with $c$-number fields. The last two lines
are discrete transformations which are not a symmetry of the classical action, but depend on the fact that the $\psi_i$'s are
Grassmann variables.  Note that charge conjugation is given by
\begin{equation}
\psi_c = \gamma_5 \psi^* = \left( \begin{array}{r} - \psi_1^* \\  \psi_2^* \end{array} \right) 
\label{B6}
\end{equation}
in our representation, so that the discrete Pauli-G\"ursey transformations can be thought of as combinations of chiral
transformations and charge conjugation. The group structure behind (\ref{B5}) is
O(2)$_R$$\otimes$O(2)$_L$, if we decompose $\psi_1,\psi_2$ into real and imaginary parts, an extension of the chiral symmetry group 
SO(2)$_R$$\otimes$SO(2)$_L$.
\section{Duality transformation of GN and NJL models}
\label{sect3}
Let us consider the original GN model (\ref{A1}) with discrete chiral symmetry first.
Its Lagrangian reads (using the summation convention for the flavor indices $i=1,...,N$) 
\begin{equation}
{\cal L}_{\rm GN}  =     - 2i \psi_1^{(i)*}\partial \psi_1^{(i)} + 2 i \psi_2^{(i)*}\bar{\partial} \psi_2^{(i)} 
+ \frac{g^2}{2}  \left( \psi_1^{(i)*}\psi_2^{(i)} +  \psi_2^{(i)*}\psi_1^{(i)} \right)^2 \,  .
\label{C1}
\end{equation}
Out of the Pauli-G\"ursey group (\ref{B5}), U(1)$_V$ (conservation of fermion number), the Z$_2$ chiral subgroup 
($\psi_1 \to \pm\psi_1, \psi_2 \to \pm \psi_2$) and charge conjugation are unbroken by the interaction term.
If we perform the canonical transformation $\psi_1 \to \psi_1^*$ which leaves only the free part of the Lagrangian invariant,
we generate a new interacting theory which will also be integrable \cite{L23}. This transformation will be referred to as ``duality transformation" from now on. The 
Lagrangian dual to (\ref{C1}) is
\begin{equation}
\tilde{\cal L}_{\rm GN}  =     - 2i \psi_1^{(i)*}\partial \psi_1^{(i)} + 2 i \psi_2^{(i)*}\bar{\partial} \psi_2^{(i)} 
+ \frac{g^2}{2}  \left( \psi_1^{(i)*}\psi_2^{(i)*} +  \psi_2^{(i)}\psi_1^{(i)} \right)^2 \,  .
\label{C2}
\end{equation}
This Lagrangian gives rise to fermion-fermion pairing instead of fermion-antifermion pairing, i.e., features superconductivity rather than chiral
symmetry breaking. The residual Pauli-G\"ursey symmetries are now U(1)$_A$ (conservation of axial charge), Z$_2$ chiral symmetry, and charge
conjugation.
The Cooper pair condensate in this model is real, as is the chiral condensate in the original GN model. We do not have to solve the superconductivity
model (\ref{C2}), but can take over all results from the massless GN model after an appropriate re-interpretation of the observables. 

Next we turn to the NJL model (\ref{A2}). Here, the continuous chiral symmetry of the Pauli-G\"ursey group is preserved, as is manifest in the chiral
representation of the Dirac matrices (\ref{B1}) where the NJL Lagrangian takes on the form 
\begin{equation}
{\cal L}_{\rm NJL} = - 2i \psi_1^{(i)*} \partial \psi_1^{(i)} + 2 i \psi_2^{(i)*} \bar{\partial} \psi_2^{(i)}
+ 2 g^2  \left( \psi_1^{(i)*}\psi_2^{(i)} \right)  \left( \psi_2^{(j)*}\psi_1^{(j)} \right) \, .
\label{C3}
\end{equation}
The discrete part of the Pauli-G\"ursey group breaks down to charge conjugation. Applying the duality transformation to this Lagrangian yields the following 
four-fermion theory,
\begin{equation}
\tilde{\cal L}_{\rm NJL} = - 2i \psi_1^{(i)*} \partial \psi_1^{(i)} + 2 i \psi_2^{(i)*} \bar{\partial} \psi_2^{(i)}
+ 2 g^2 \left( \psi_1^{(i)*}\psi_2^{(i)*} \right)  \left( \psi_2^{(j)}\psi_1^{(j)} \right) \, .
\label{C4}
\end{equation}
This is yet another field theory with fermion-fermion pairing, here with the full U(1)$_V$ and U(1)$_A$ symmetries.
As pointed out in Ref.~\cite{L23}, it is in fact identical to the Cooper pair Lagrangian proposed by Chodos, Minakata and Cooper (CMC) \cite{L7},
\begin{equation}
 {\cal L}_{\rm CMC} = \bar{\psi}^{(i)} i\partial \!\!\!/ \psi^{(i)}
+ 2 G^2 \left( \bar{\psi}^{(i)} \gamma_5 \psi^{(j)}\right)  \left(\bar{\psi}^{(i)} \gamma_5 \psi^{(j)}\right)  \, ,
\label{C5}
\end{equation}
for the choice $g^2=2 G^2$. Since the duality transformation is a canonical transformation, there is no need to solve the Cooper pair model anew if the NJL
model has been solved already. All one has to do is translate the physical observables into the dual language. Thus for instance, the chiral vacuum circle of the NJL model
becomes the circle of the complex Cooper pair condensates in the dual model. Integrability of the Cooper pair model (\ref{C4},\ref{C5}) again follows
trivially from that of the NJL model. As also discussed in Ref.~\cite{L23}, the situation becomes slightly more involved if one includes vector
and axial chemical potentials $\mu, \mu_5$. Since the vector and axial vector densities change their role under the duality transformation $\psi_1 \to \psi_1^*$,
one has to interchange $\mu$ and $\mu_5$, see also Ref.~\cite{L12}. 

The content of this section is in essence already contained in Ref.~\cite{L23}. The reason why we recall it here is to set the stage for more interesting
candidates of integrable models. They have both fermion-fermion and fermion-antifermion pairing and will be discussed in the following two sections.

\section{Self-dual GN model}
\label{sect4}
The GN model (\ref{C1}) breaks the discrete part of the Pauli-G\"ursey group  down to charge conjugation. This enabled us to generate a new integrable 
model by applying the duality transformation ($\psi_1 \to \psi_1^*$) to the GN Lagrangian, see Eq.~(\ref{C2}). We now try to construct another integrable model by ``self-dualizing"
the GN Lagrangian. This means that we add the interaction term of the dual GN model (\ref{C2}) to the GN model Lagrangian (\ref{C1}), so that the full Lagrangian
shares the discrete part of the Pauli-G\"ursey symmetry with the free, massless theory,
\begin{eqnarray}
{\cal L}_{\rm sdGN} & = &    - 2i \psi_1^{(i)*}\partial \psi_1^{(i)} + 2 i \psi_2^{(i)*}\bar{\partial} \psi_2^{(i)} 
\nonumber \\
& & +  \frac{g^2}{2} \left[ \left( \psi_1^{(i)*}\psi_2^{(i)} +  \psi_2^{(i)*}\psi_1^{(i)} \right)^2 
+  \left( \psi_1^{(i)*}\psi_2^{(i)*} +  \psi_2^{(i)}\psi_1^{(i)} \right)^2 \right] \, .
\label{D1}
\end{eqnarray}
We shall refer to this model as self-dual Gross-Neveu (sdGN) model. Notice that this theory possesses neither U(1)$_V$ nor U(1)$_A$ symmetries. The discrete 
symmetry is recognized as the dihedral group D$_2$ (the symmetry group of a rectangle) with four elements ($\psi_1 \to \pm \psi_1, \pm \psi_1^*$). The two
interaction terms in (\ref{D1}) can give rise to both fermion-fermion and fermion-antifermion pairing, both condensates being real.

In view of the large $N$ limit, we perform a standard Hubbard-Stratonovich transformation \cite{L27,L28} on the Lagrangian (\ref{D1}).
The Lagrangian  
\begin{eqnarray}
{\cal L}'_{\rm sdGN} & = &   {\cal L}_{\rm sdGN}  - \frac{1}{2g^2} \left[ {\cal S} + g^2 \left(  \psi_1^{(i)*} \psi_2^{(i)} + \psi_2^{(i)*} \psi_1^{(i)} \right)   \right]^2
\nonumber \\
& & -  \frac{1}{2g^2} \left[ {\cal B} + g^2 \left(  \psi_1^{(i)*} \psi_2^{(i)*} + \psi_2^{(i)} \psi_1^{(i)} \right)   \right]^2 \, .
\label{D2}
\end{eqnarray}
involving two real, scalar, flavor singlet fields ${\cal S}, {\cal B}$
is equivalent to ${\cal L}_{\rm sdGN}$ from Eq.~(\ref{D1}). It can be expanded as 
\begin{eqnarray}
{\cal L}'_{\rm sdGN} & = & - 2i \psi_1^{(i)*}\partial \psi_1^{(i)} + 2 i \psi_2^{(i)*}\bar{\partial} \psi_2^{(i)}  - {\cal S}   \left( \psi_1^{(i)*}\psi_2^{(i)} +  \psi_2^{(i)*}\psi_1^{(i)} \right) 
\nonumber \\
& & - {\cal B} \left(   \psi_1^{(i)*}\psi_2^{(i)*} + \psi_2^{(i)}\psi_1^{(i)}  \right) - \frac{1}{2g^2} \left( {\cal S}^2+{\cal B}^2 \right)\, .
\label{D3}
\end{eqnarray}
The Euler-Lagrange equations for ${\cal S}, {\cal B}$ show that these fields are constrained  to fermion bilinears as follows,
\begin{eqnarray}
{\cal S} & = & -  g^2 \left( \psi_1^{(i)*} \psi_2^{(i)} + \psi_2^{(i)*} \psi_1^{(i)} \right) \, ,
\nonumber \\
{\cal B} & = & -  g^2  \left( \psi_1^{(i)*} \psi_2^{(i)*} +  \psi_2^{(i)} \psi_1^{(i)} \right) \, .
\label{D4}
\end{eqnarray}
In the large $N$ limit, the auxiliary fields can be replaced by their expectation values, so that one has to deal 
with quantized fermions in a $c$-number mean field to be determined self-consistently,
according to the expectation values of (\ref{D4}). For the standard GN models with fermion-antifermion
pairing, these last few steps can be regarded as derivation of the relativistic Hartree-Fock (HF) framework. Similarly, the
more general case with fermion-fermion and fermion-antifermion pairing leads to the 
Hartree-Fock-Bogoliubov (HFB) theory, a generalization of HF well known in many-body physics \cite{L26}.

In order to quantize the theory canonically, we turn to the Hamiltonian density corresponding to ${\cal L}'_{\rm sdGN}$ 
\begin{eqnarray}
{\cal H} & = &  i \psi_1^{(i)*}  \partial_1 \psi_1^{(i)} - i \psi_2^{(i)*} \partial_1 \psi_2^{(i)} 
+ {\cal S}   \left( \psi_1^{(i)*}\psi_2^{(i)}  +  \psi_2^{(i)*}\psi_1^{(i)} \right) 
\nonumber \\
& & + {\cal B} \left(  \psi_1^{(i)*}\psi_2^{(i)*} + \psi_2^{(i)}\psi_1^{(i)}   \right) 
 + \frac{1}{2g^2} \left( {\cal S}^2 + {\cal B}^2 \right) \, .
\label{D5}
\end{eqnarray}
Keeping only the fermionic part for the moment, we can cast the quantized Hamiltonian into the Nambu-Gorkov form \cite{L29,L30}, 
\begin{equation}
H = \frac{1}{2} \int dx \left( \psi_1^{\dagger}, \psi_2^{\dagger}, \psi_1, \psi_2  \right) 
\left( \begin{array}{cccc} i \partial_1 & {\cal S} & 0 & {\cal B} \\ {\cal S} &  -i \partial_1 & -{\cal B} & 0 \\
0 & -{\cal B}  & i\partial_1  & -{\cal S} \\ {\cal B} & 0 & -{\cal S} & -i\partial_1 \end{array} \right)
  \left( \begin{array}{c} \psi_1 \\ \psi_2  \\ \psi_1^{\dagger} \\ \psi_2^{\dagger} \end{array} \right)\, ,
\label{D6}
\end{equation}
where we have suppressed flavor indices.
The 4$\times$4 matrix appearing in (\ref{D6}) will be denoted by $h$ from now on. It plays the role of the first-quantized HFB
Hamiltonian. We first note an important symmetry property which will be useful later on, invariance under charge conjugation.
According to (\ref{B6}), charge conjugation 
in the space of Nambu-Gorkov spinors can be represented through a unitary matrix,
\begin{equation}
\left( \begin{array}{c} \psi^{\dagger} \\ \psi \end{array} \right)_c = \left( \begin{array}{cc} 0 & \gamma_5 \\ \gamma_5 & 0 \end{array} 
\right) \left( \begin{array}{c} \psi^{\dagger} \\ \psi \end{array} \right)\, .
\label{D7}
\end{equation}
Denoting this unitary matrix by $U_c$, it is easy to verify that
\begin{equation}
h = U_c h U_c^{\dagger} \, .
\label{D7a}
\end{equation}
Next we observe that $h$ can be block-diagonalized by a constant, unitary transformation $V$,
\begin{equation}
h  =  V^{\dagger} h_{\rm bd} V, \quad h_{\rm bd} = \left( \begin{array}{cc} h_I & 0 \\ 0 & h_{II} \end{array} \right) \, ,
\label{D8}
\end{equation}
with
\begin{equation}
V=\frac{1}{\sqrt{2}} \left( \begin{array}{rrrr} 1 & 0 & 1 & 0 \\ 0 & 1 & 0 & -1 \\ 1 & 0 & -1 & 0 \\ 0 & 1 & 0 & 1 \end{array} \right), \quad VV^{\dagger}=1 ,
\quad h_{I,II}= \left( \begin{array}{cc} i \partial_1 & S_{I,II} \\ S_{I,II} & -i\partial_1 \end{array} \right),
\label{D9}
\end{equation}
and $S_I = {\cal S}-{\cal B}, S_{II}={\cal S}+{\cal B}$.
Note that each 2$\times$2 block $h_{I,II}$ looks like the first quantized Hamiltonian of the standard GN model in HF approximation 
with scalar mean fields ${\cal S} \mp {\cal B}$.
In order to simplify further the HFB Hamiltonian, we plug (\ref{D8}) into (\ref{D6}),
\begin{equation}
H = \frac{1}{2} \int dx \left( \psi^{\dagger}, \psi \right) V^{\dagger} h_{\rm bd} V \left( \begin{array}{c} \psi \\ \psi^{\dagger} \end{array} \right)
= \frac{1}{2} \int dx \Psi^{\dagger} h_{\rm bd} \Psi \, .
\label{D10}
\end{equation}
In the last step we have introduced unitarily transformed fermion field operators 
\begin{equation}
\Psi = V \left( \begin{array}{c} \psi \\ \psi^{\dagger} \end{array} \right) = \frac{1}{\sqrt{2}} \left( \begin{array}{c} \psi_1+ \psi_1^{\dagger} \\  \psi_2- \psi_2^{\dagger} \\ 
 \psi_1- \psi_1^{\dagger} \\ 
 \psi_2+ \psi_2^{\dagger} \end{array} \right) := \left( \begin{array}{r} \chi_1 \\ i \chi_2 \\ -i \chi_3 \\ \chi_4 \end{array} \right)  \, .
\label{D11}
\end{equation}
Thus block-diagonalization of the Hamiltonian matrix $h$ reveals that the natural degrees of freedom are four independent Majorana fields per flavor
($\chi_a^{\dagger}=\chi_a$) obeying the anticommutation relations 
\begin{equation}
  \{ \chi_a^{(i)}(x), \chi_b^{(j)}(y) \}  = \delta_{ab}\delta_{ij} \delta(x-y) \, .
\label{D12}
\end{equation}
(The choice of signs in the definition of the $\chi_a$'s is a matter of convention and was made in such a way that some formulas below simplify.)
In these variables, the HFB Hamiltonian of the self-dual GN model decomposes into a sum of two commuting Hamiltonians,
\begin{equation}
H =  H_I + H_{II} \, ,
\label{D13}
\end{equation}
with
\begin{eqnarray}
H_I & = & \frac{1}{2} \int dx \left( \chi_1, \chi_2 \right) \left( \begin{array}{cc} i\partial_1 & iS_I \\ -iS_I & -i \partial_1 \end{array} \right)
\left( \begin{array}{c} \chi_1 \\ \chi_2 \end{array} \right) \, ,
\nonumber \\
H_{II} & = & \frac{1}{2} \int dx \left( \chi_3, \chi_4 \right) \left( \begin{array}{cc} i\partial_1 & iS_{II} \\ -iS_{II} & -i \partial_1 \end{array} \right)
\left( \begin{array}{c} \chi_3 \\ \chi_4 \end{array} \right) \, .
\label{D14}
\end{eqnarray}
In principle, these two terms could still be coupled via the scalar fields $S_{I,II}$ through the self-consistency condition in the HFB approach,
but we will now verify that this is not the case. 
The inverse relations to (\ref{D11}),
\begin{eqnarray}
\psi_1 & = & \frac{1}{\sqrt{2}} \left( \chi_1 - i \chi_3 \right) \, ,
\nonumber \\
\psi_2 & = & \frac{1}{\sqrt{2}} \left( \chi_4 + i \chi_2 \right) \, ,
\label{D15}
\end{eqnarray}
can be used to translate ${\cal S},{\cal B}$ in Eq.~(\ref{D4}) into Majorana fields,
\begin{eqnarray}
{\cal S} & = & -ig^2 \left( \chi_3^{(i)} \chi_4^{(i)} +  \chi_1^{(i)} \chi_2^{(i)} \right) \, ,
\nonumber \\
{\cal B} & = & - ig^2 \left( \chi_3^{(i)} \chi_4^{(i)} -  \chi_1^{(i)} \chi_2^{(i)} \right) \, .
\label{D16}
\end{eqnarray}
Hence, in the large $N$ limit,
\begin{eqnarray}
S_I & = & -2ig^2 \langle  \chi_1^{(i)} \chi_2^{(i)} \rangle \, ,
\nonumber \\
S_{II} & = & -2ig^2 \langle \chi_3^{(i)} \chi_4^{(i)} \rangle \, ,
\label{D17}
\end{eqnarray}
so that the full HFB problem indeed separates into two simpler, independent problems of HF type. As a matter of fact, $H_{I,II}$ and the self-consistency
conditions (\ref{D17}) are the same as in the standard GN model, but with Majorana instead of Dirac fields (the O($N$) symmetric model, rather than the 
U($N$) or O($2N$) symmetric model with Dirac fermions). This shows at once that the sdGN model is integrable, and that its solution can be reduced
to solutions of the standard GN model. 

One point still has to be clarified: We have dropped the purely bosonic part from the Hamiltonian, which contains the coupling constant $g^2$ of the
sdGN model. This coupling constant does not have to coincide with $G^2$, the one of the pair of standard GN models. 
We shall determine $G^2$ by demanding that the bosonic part of the Hamiltonian be also additive,
\begin{equation}
\frac{{\cal S}^2 + {\cal B}^2}{2g^2} =  \frac{S_I^2  +S_{II}^2}{2G^2}, \quad  \quad S_{I,II}={\cal S} \mp {\cal B} \, .
\label{D18}
\end{equation}
This fixes the GN coupling constant to the value $G^2=2g^2$. We will confirm this choice via the self-consistency conditions
of the GN and sdGN models below.

We were led to introduce Majorana fields as a result of 
block-diagonalization of the HFB Hamiltonian $h$. To better understand this result, let us go back to the Lagrangian (\ref{D1}) and express the 
Dirac fields in terms of Majorana fields right away, using (\ref{D15}),
\begin{eqnarray}
{\cal L}_{\rm sdGN} & = & - i \chi_1^{(i)} \partial \chi_1^{(i)} + i \chi_2^{(i)} \bar{\partial} \chi_2^{(i)} - g^2 \left( \chi_1^{(i)} \chi_2^{(i)}\right)^2
\nonumber \\
& & - i \chi_3^{(i)} \partial \chi_3^{(i)} + i \chi_4^{(i)} \bar{\partial} \chi_4^{(i)} - g^2 \left( \chi_3^{(i)} \chi_4^{(i)}\right)^2 \, .
\label{D19}
\end{eqnarray}
This is indeed a sum of two independent O($N$) GN Lagrangians. Although we arrived at our  findings through the HFB  approach, this simple exercise
shows that they have nothing to do with it, but can be exposed already at the level of the Lagrangian.

The O($N$) symmetric GN model with $N$ Majorana fields is equivalent to the 
U($N/2$) symmetric GN model with $N/2$ Dirac fields. Since the solutions of the GN model are usually formulated for Dirac fields, we first transform expressions
(\ref{D14},\ref{D17}) into Dirac language. There are many ways how to combine pairs of Majorana fields into Dirac fields, due to the flavor degrees of 
freedom. One possible choice is
\begin{eqnarray}
\psi_{I,1}^{(i)} = \frac{1}{\sqrt{2}}\left( \chi_1^{(i)}-i \chi_1^{(N/2+i)} \right) \, ,
\nonumber \\
\psi_{I,2}^{(i)} = \frac{1}{\sqrt{2}}\left( \chi_2^{(N/2+i)}+i \chi_2^{(i)} \right) \, ,
\nonumber \\
\psi_{II,1}^{(i)} = \frac{1}{\sqrt{2}}\left( \chi_3^{(i)}-i \chi_3^{(N/2+i)} \right) \, ,
\nonumber \\
\psi_{II,2}^{(i)} = \frac{1}{\sqrt{2}}\left( \chi_4^{(N/2+i)}+i \chi_4^{(i)} \right) \, ,
\label{D20}
\end{eqnarray}
for $i=1,...,N/2$. This yields
\begin{equation}
H_{I}  =  \int dx \sum_{i=1}^{N/2} \left( \psi_{I,1}^{(i)\dagger}, \psi_{I,2}^{(i)\dagger} \right) \left( \begin{array}{cc} i \partial_1 & S_{I} \\ S_{I} & -i \partial_1 \end{array} \right)
\left( \begin{array}{c} \psi_{I,1}^{(i)} \\ \psi_{I,2}^{(i)} \end{array} \right) \, ,
\label{D21}
\end{equation}
and a similar equation with all subscripts $I$ replaced by $II$.
The condensate operators, assuming that the two standard GN models have coupling constant $G^2$, read
\begin{eqnarray}
S_I & = & - G^2 \sum_{i=1}^{N/2} \left( \psi_{I,1}^{(i)\dagger} \psi_{I,2}^{(i)} +  \psi_{I,2}^{(i)\dagger} \psi_{I,1}^{(i)} \right) = -i G^2 \chi_1^{(i)} \chi_2^{(i)} \, ,
\nonumber \\
S_{II} & = & - G^2 \sum_{i=1}^{N/2} \left( \psi_{II,1}^{(i)\dagger} \psi_{II,2}^{(i)} +  \psi_{II,2}^{(i)\dagger} \psi_{II,1}^{(i)} \right) = - i G^2 \chi_3^{(i)} \chi_4^{(i)}  \, ,
\label{D22}
\end{eqnarray}
where we continue to use the summation convention for indices running from 1 to $N$.
This agrees with (\ref{D17}) provided we set $G^2=2 g^2$, confirming our findings from the bosonic part of the Hamiltonian. Notice that in this case the 't~Hooft condition reads
\begin{equation}
\frac{N}{2} G^2 = N g^2 = {\rm const.}
\label{D23}
\end{equation}
Thus the sdGN model with $N$ Dirac flavors and coupling constant $g^2$ is mapped onto two independent GN models with $N/2$ Dirac flavors each and coupling constant
$2g^2$. The value of the 't~Hooft coupling, $Ng^2$, is the same in the sdGN model and the two GN models.

We will now show how to construct a self-consistent HFB solution of the sdGN model out of any pair of self-consistent HF solutions of the standard GN model. 
To this end we immediately go the time-dependent version of HF and HFB theory, since this is not more complicated than the static case. The time dependent Hartree-Fock equations
(TDHF) for the two independent GN models can be cast into the form
\begin{equation}
\left( \begin{array}{cccc} 2 i \partial & S_I & 0 & 0 \\
S_I & - 2 i \bar{\partial} & 0 & 0 \\ 0 & 0 & 2 i \partial &  S_{II} \\ 0 & 0 & S_{II} & - 2i \bar{\partial} \end{array} \right)  \left( \begin{array}{c} \varphi_{I,1} \\ \varphi_{I,2} \\
\varphi_{II,1} \\ \varphi_{II,2} \end{array} \right) = 0 \, .
\label{D24}
\end{equation}
Here, the spinors are solutions of the Dirac equation describing the single particle levels. They have to fulfill the self-consistency conditions
\begin{eqnarray}
S_{I} & = &  - Ng^2 \sum^{\rm occ} \left( \varphi_{I,1}^*\varphi_{I,2} + \varphi_{I,2}^* \varphi_{I,1} \right) \, ,
\nonumber \\
S_{II} & = &  - Ng^2 \sum^{\rm occ} \left( \varphi_{II,1}^*\varphi_{II,2} + \varphi_{II,2}^* \varphi_{II,1} \right) \, ,
\label{D25}
\end{eqnarray}
where the sum runs over all occupied states. The time-dependent Hartree-Fock-Bogoliubov (TDHFB) equation for the sdGN model 
on the other hand can be written as the following system of four coupled equations,
\begin{equation}
\left( \begin{array}{cccc} 2 i \partial & {\cal S} & 0 & {\cal B} \\ {\cal S} & - 2 i \bar{\partial} & -{\cal B} & 0 \\ 0 & -{\cal B} & 2i\partial & -{\cal S} \\
{\cal B} & 0 & -{\cal S} & - 2i \bar{\partial} \end{array} \right) \left( \begin{array}{c} \phi_{1} \\ \phi_{2} \\
\phi_{3} \\ \phi_{4} \end{array} \right) = 0 \, , 
\label{D26}
\end{equation}
supplemented by the self-consistency conditions
\begin{eqnarray}
{\cal S} & = & - \frac{Ng^2}{2} \sum^{\rm occ} \left( \phi_1^* \phi_2 + \phi_2^*\phi_1 - \phi_4^* \phi_3 - \phi_3^* \phi_4 \right)\, ,
\nonumber \\
{\cal B} & = &  -\frac{Ng^2}{2} \sum^{\rm occ} \left( \phi_1^* \phi_4 + \phi_4^*\phi_1 - \phi_2^* \phi_3 - \phi_3^* \phi_2 \right)\, .
\label{D27}
\end{eqnarray}
It is now easy to verify that the unitary transformation $V$, Eq.~(\ref{D9}), transforms equations (\ref{D26}) into (\ref{D24}) and the self-consistency condition 
(\ref{D27}) into (\ref{D25}), remembering that $S_{I,II} = {\cal S} \mp {\cal B}$. Hence the Nambu-Gorkov spinors for any single quasi-particle level of the 
sdGN model are related to the GN spinors via
\begin{equation}
\left( \begin{array}{c} \phi_{1} \\ \phi_{2} \\ \phi_{3} \\ \phi_{4} \end{array} \right) 
= V^{\dagger} \left( \begin{array}{c} \varphi_{I,1} \\ \varphi_{I,2} \\ \varphi_{II,1} \\ \varphi_{II,2} \end{array} \right) = \frac{1}{\sqrt{2}}
\left( \begin{array}{c} \varphi_{I,1} + \varphi_{II,1} \\ \varphi_{I,2} + \varphi_{II,2} \\ \varphi_{I,1}-\varphi_{II,1} \\ \varphi_{II,2}-\varphi_{I,2} \end{array} \right) \, . 
\label{D28}
\end{equation}
This looks at first sight as if one would have to add and subtract spinors from two different GN solutions. However, this is not the case. The correct interpretation
of Eq.~(\ref{D28}) is as follows. If we take any solution of the GN model labelled $I$ in (\ref{D24}), we set $\varphi_{II,1} = \varphi_{II,2} = 0$ and obtain the 
sdGN model spinors
\begin{equation}
\Phi_{I} = \frac{1}{\sqrt{2}} \left( \begin{array}{r} \varphi_{I,1} \\ \varphi_{I,2} \\
\varphi_{I,1} \\ -\varphi_{I,2} \end{array} \right)\, .
\label{D28a}
\end{equation} 
The contribution from the occupied states to the mean field ${\cal S}$ ($\cal B$) is $S_I/2$ ($-S_I/2$), see (\ref{D25},\ref{D27}). The GN model labelled $II$
corresponds to setting $\varphi_{I,1}=\varphi_{I,2}=0$ in (\ref{D24}) and consequently to the sdGN spinors
\begin{equation}
\Phi_{II} = \frac{1}{\sqrt{2}} \left( \begin{array}{r} \varphi_{II,1} \\ \varphi_{II,2} \\
- \varphi_{II,1} \\ \varphi_{II,2} \end{array} \right)\, .
\label{D28b}
\end{equation} 
Their contribution to both ${\cal S}$ and ${\cal B}$ is $S_{II}/2$, so that the relations $S_{I,II} = {\cal S} \mp {\cal B}$ are indeed satisfied. Notice also that the
quasi-particle spinors $\Phi_{I,II}$ are eigenstates of the charge conjugation matrix $U_c$ from Eq.~(\ref{D7}),
\begin{equation}
U_c \Phi_I = - \Phi_I, \quad U_c \Phi_{II} = \Phi_{II}\, .
\label{D28c}
\end{equation}
We will come back to this observation below when we interpret the two decoupled GN models in more physical terms.

This shows that the TDHFB solution of the self-dual GN model inherits self-consistency from the two
input solutions of the standard GN model. The energy is the sum of the energies of both constituent solutions, since this also holds for the 
Hamiltonians, as discussed above. Since the massless GN model is integrable and its complete large-$N$ solution is known analytically, the same is 
true for the self-dual variant of the GN model. 

In the GN model, the Z$_2$ chiral symmetry maps the TDHF solution with condensate $S$ onto the one with condensate $-S$ after spontaneous symmetry breaking (SSB).
In the self-dual GN model, the Pauli-G\"ursey D$_2$ symmetry
maps the TDHFB solution with condensates $({\cal S},{\cal B})$ onto the ones with $(-{\cal S},-{\cal B})$, $({\cal B},{\cal S})$, $(-{\cal B}, -{\cal S})$. The change of sign is due to
the discrete chiral symmetry, whereas swapping ${\cal S}$ and ${\cal B}$ is the result of the duality transformation.

Clearly, one can generate a huge variety of static and dynamical solutions of the TDHFB equation in this way. Let us explore 
some of the simpler solutions to see whether they make sense from the physics point of view.

\vskip 0.2cm
{\bf 1) Vacua}
\vskip 0.2cm
The GN model with spontaneously broken Z$_2$ chiral symmetry has two degenerate vacua with $S=\pm m = \pm 1$ (dynamical fermion mass in natural units), see e.g. \cite{L31}.
Consequently there are four degenerate vacua
in the self-dual GN model, see Fig. 1, reflecting SSB of the larger discrete group D$_2$. The ground state is either a superconductor (${\cal S}=0, {\cal B}=\pm 1$) or a 
chirally broken state (${\cal S}=\pm 1, {\cal B}=0$). All four states are physically indistinguishable, as they differ only in the convention for the fermion operators.
The renormalized vacuum energy density is the same as in the GN model with $N$ Dirac flavors, $-N/4\pi$.
\begin{figure}[h]
\begin{center}
\epsfig{file=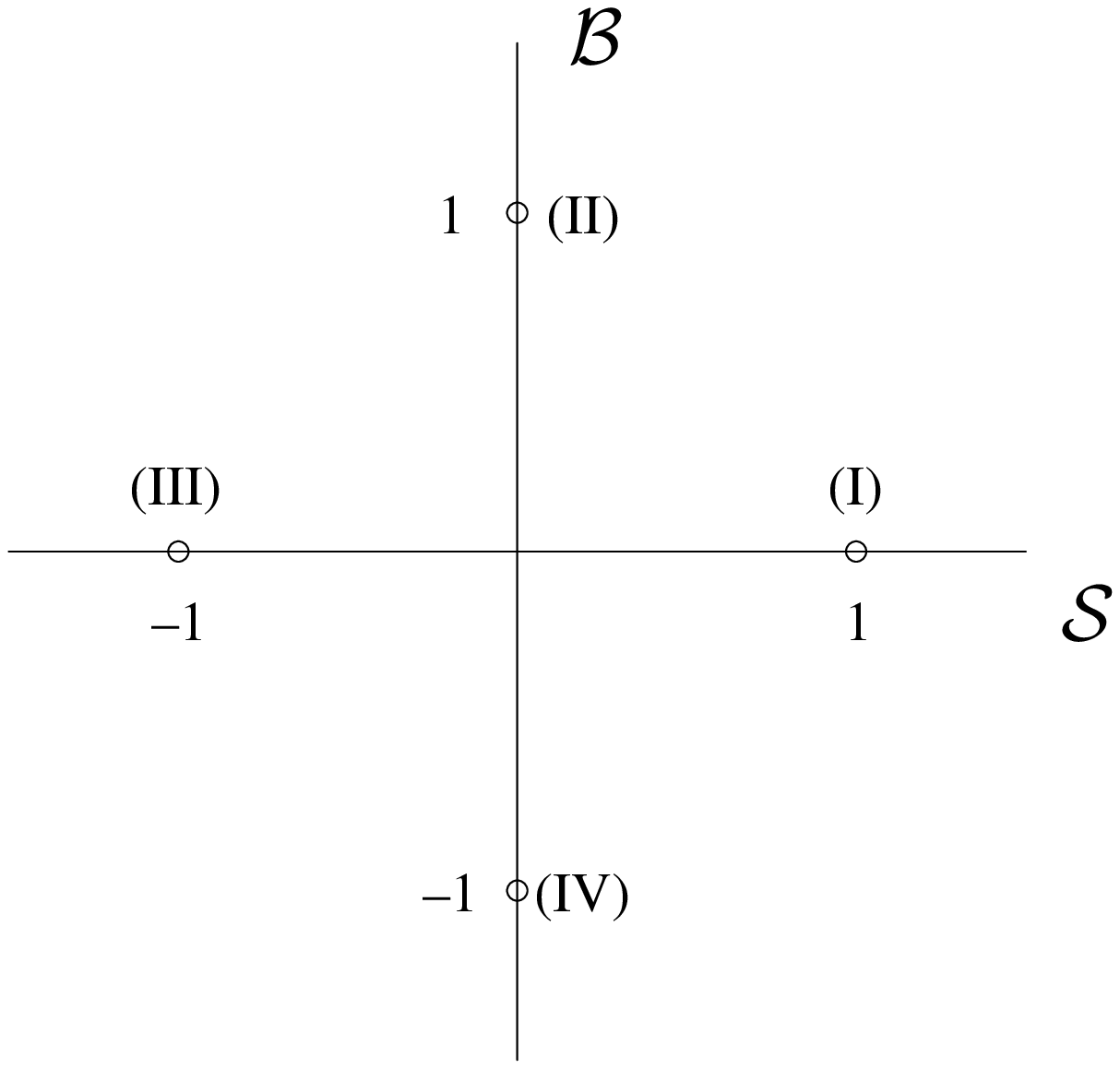,width=6cm,height=6cm,angle=270}
\caption{Four vacua of sdGN model.}
\label{Fig1}
\end{center}
\end{figure}

\vskip 0.2cm
{\bf 2) Kinks} 
\vskip 0.2cm
In the GN model, the kink interpolates between the two vacua with $S=\pm 1$ \cite{L31}. In the self-dual GN model we expect six types of ``domain walls" 
separating two out of the four vacua. They can easily be found by using as input either the GN kink and the vacuum, or two GN kinks.
If we choose the vacuum and a kink for $S_{\pm}$,
we get the kinks between two neighboring vacua (I and II, II and III, III and IV, IV and I), see Fig. 2, whose mass is half of the mass of a GN kink with $N$ Dirac flavors,
$N/2\pi$.
\begin{figure}[h]
\begin{center}
\epsfig{file=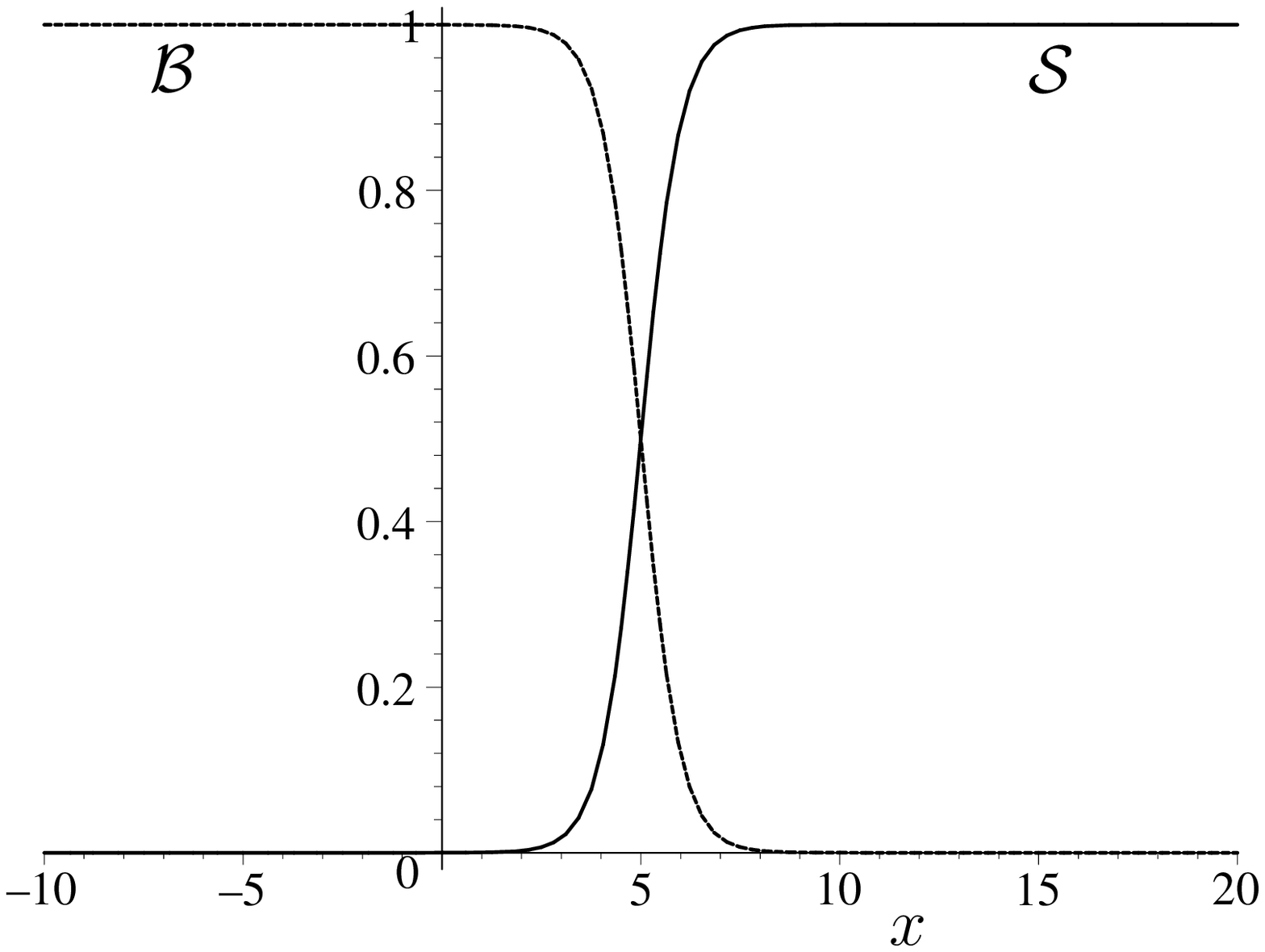,width=6cm,height=6cm,angle=270}
\caption{Static kink joining vacua II and I of the sdGN model, composed of GN vacuum and GN kink. Formulas in Eq.~(\ref{D29}).}
\label{Fig2}
\end{center}
\end{figure}
\begin{figure}[h]
\begin{center}
\epsfig{file=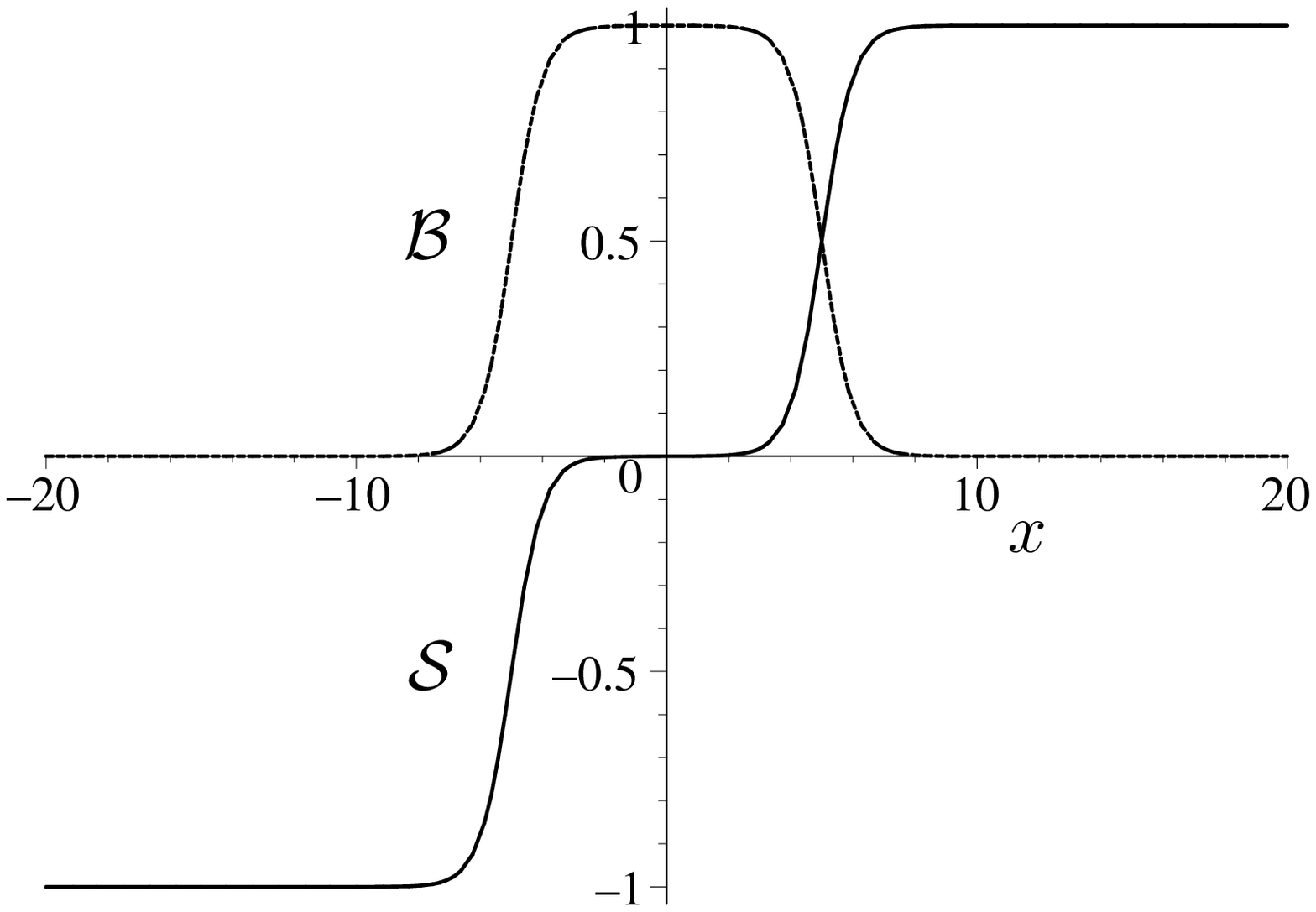,width=6cm,height=6cm,angle=270}
\caption{Static kink joining vacua III and I of the sdGN model, composed of two (shifted) GN kinks. Formulas in Eq.~(\ref{D30}), $a=-5$ shown.}
\label{Fig3}
\end{center}
\end{figure}
If we choose two kinks which are shifted relative to 
each other, we interpolate between two opposite vacua (I and III, II and IV), see Fig. 3. Here, the mass is equal to the mass of the GN kink with $N$ Dirac flavors, $N/\pi$. 
The width of this kink can be made arbitraryly large by pulling the constituent kinks apart. In the transition region, there is a localized zone where the system
is in the dual vacuum. This can be used for instance to manufacture a domain wall between the ${\cal B}=1$ and ${\cal B}=-1$ superconducting vacua,
separated by a normal (chirally broken) region --- a kind of $\pi$-Josephson junction (the dual of Fig. 3).
We list the expressions for the condensates for the examples of kinks shown in Fig. 2,
\begin{equation}
{\cal S}  =  \frac{1}{2} \left( 1+\tanh x \right), \quad {\cal B} = \frac{1}{2} \left( 1- \tanh x \right) \, ,
\label{D29}
\end{equation}
and in Fig. 3,
\begin{equation}
{\cal S}  =  \frac{1}{2} \left[ \tanh (x-a) +\tanh (x+a) \right], \quad {\cal B} = \frac{1}{2} \left[ \tanh (x-a) - \tanh (x+a) \right] \, .
\label{D30}
\end{equation}
In the first case, if one plots the corresponding trajectories in the (${\cal S},{\cal B}$) plane, one gets a straight line segment from vacuum II to vacuum I.
In the second case the result depends on the shift parameter $a$, see Fig. 4. 
\begin{figure}[h]
\begin{center}
\epsfig{file=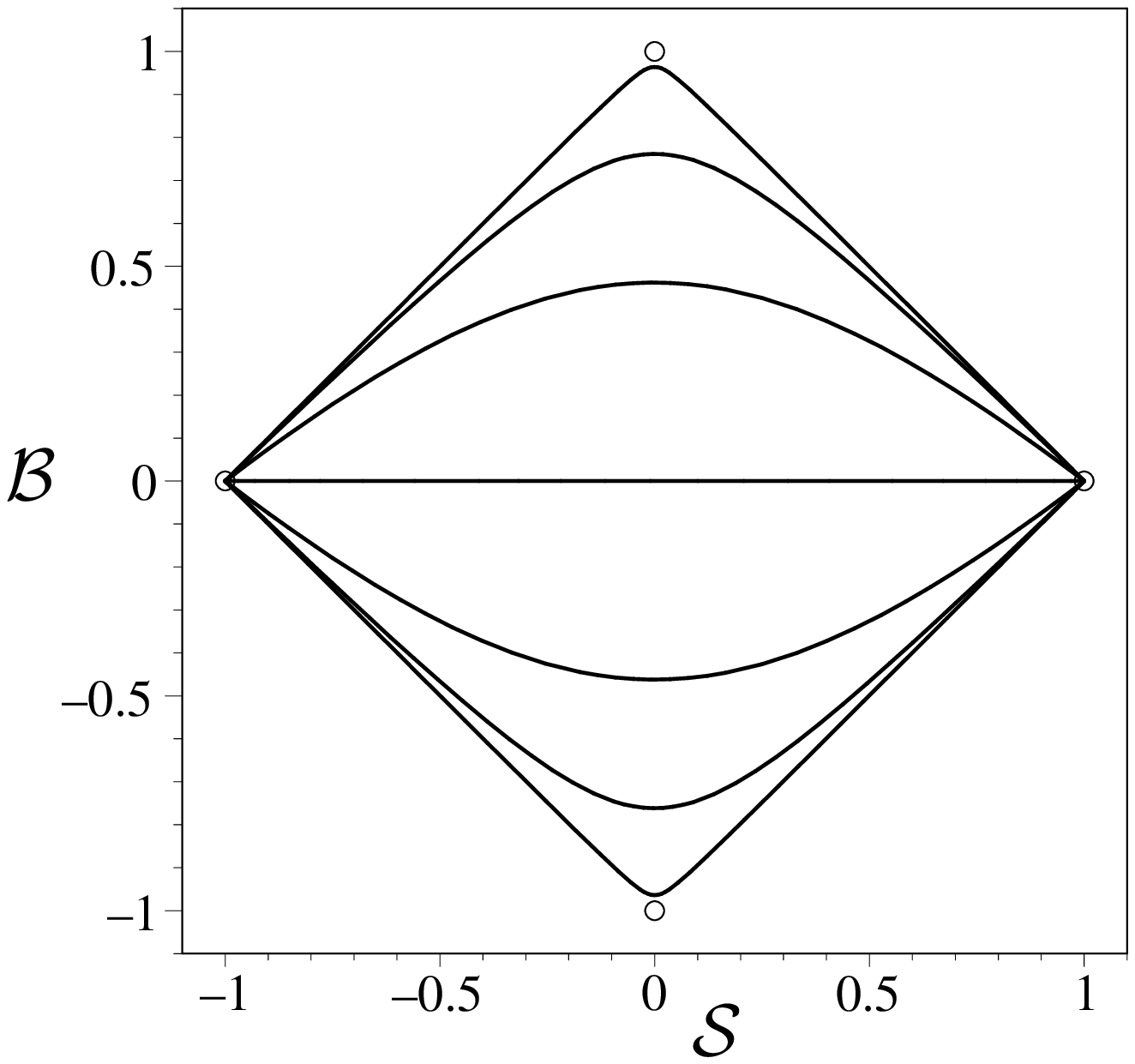,width=6cm,height=6cm,angle=270}
\caption{(${\cal S},{\cal B}$)-plot of static kink joining vacua III and I of the sdGN model. Formulas in Eq.~(\ref{D30}). From top to bottom: $a=-2,-1,-0.5,0,0.5,1,2$.}
\label{Fig4}
\end{center}
\end{figure}

\vskip 0.2cm
{\bf 3) Multi-kink solutions} 
\vskip 0.2cm
Dynamical solutions result if we choose time-dependent kink solutions of the GN model as ingredients for $S_{\pm}$ \cite{L32}. A snapshot of such
a solution may be described as an arbitrary succession of regions of vacua I -- IV, separated by the kind of kinks described before (Figs. 2 and 3).
Under time evolution, these domain walls move and collide, the details depending on the input parameters. The only static kink solutions are the single
domain walls, like in the GN model.

\vskip 0.2cm
{\bf 4) Other solutions} 
\vskip 0.2cm
If we allow for DHN baryons, breathers and multi-soliton bound states, there are evidently many more static or time-dependent solutions to be explored \cite{L20,L21,L22,L33,L34,L35} 
\vskip 0.2cm
Although solitons of the sdGN model and of the GN model are so closely related, the physics interpretation may be quite different. Thus
for instance, solitons of the GN model are characterized not only by the scalar condensate, but also by the fermion density and 
fermion number they carry. In the sdGN model, neither fermion number nor axial charge are conserved, so that it is impossible 
to attribute a definite fermion number to any soliton solution. The fermion numbers of the two standard GN models is of course conserved,
but acquire a different physical meaning in the sdGN model. In fact, there are two dynamical U(1) symmetries hiding in the Lagrangian
(\ref{D1}), namely the U(1)$_V$ symmetries of the pair of equivalent GN models. 

How do these continuous symmetries of the GN models manifest themselves in the sdGN model, and what are the 
corresponding conserved Noether charges? The easiest way to answer these questions is to start from the TDHF equations (\ref{D24},\ref{D25})
and the TDHFB equations (\ref{D26},\ref{D27}), since the conservation of GN fermion number is valid separately in every single-particle state.
In the GN model, the two U(1)$_V$ transformations
read 
\begin{eqnarray}
U(1)_{V,I}: \quad \left( \begin{array}{c} \varphi_{I,1} \\ \varphi_{I,2} \end{array} \right) & \to &  e^{i \alpha} \left( \begin{array}{c} \varphi_{I,1} \\ \varphi_{I,2} \end{array} \right) \, ,
\nonumber \\
U(1)_{V,II}: \quad \left( \begin{array}{c} \varphi_{II,1} \\ \varphi_{II,2} \end{array} \right) & \to &  e^{i \beta} \left( \begin{array}{c} \varphi_{II,1} \\ \varphi_{II,2} \end{array} \right) \, .
\label{D32}
\end{eqnarray}
The corresponding Noether charges are the fermion numbers in both GN models, 
\begin{eqnarray}
Q_{I} & = &  \int dx \left( |\varphi_{I,1}|^2+ |\varphi_{I,2}|^2 \right)\, ,
\nonumber \\
Q_{II} & = &  \int dx \left( |\varphi_{II,1}|^2+ |\varphi_{II,2}|^2 \right)\, .
\label{D33}
\end{eqnarray}
(For this purpose, we assume a box normalization such that all spinors are square integrable.)
Upon using (\ref{D28}), Eq.~(\ref{D32}) translates into the following symmetry transformation of the sdGN model quasi-particle spinors,
\begin{equation}
U(1)_{V,I} \otimes U(1)_{V,II}: \quad  \left( \begin{array}{c} \phi_1 \\ \phi_2 \\ \phi_3 \\ \phi_4 \end{array} \right) \to 
 \frac{e^{i\beta} + e^{i\alpha}}{2} \left( \begin{array}{c} \phi_1 \\ \phi_2 \\ \phi_3 \\ \phi_4 \end{array} \right)
+ \frac{e^{i\beta}-e^{i\alpha}}{2}  \left( \begin{array}{r} -\phi_3 \\ \phi_4 \\ -\phi_1 \\ \phi_2 \end{array} \right)\, .
\label{D34}
\end{equation}
Remembering the charge conjugation matrix $U_c$ from Eq.~(\ref{D7}) and denoting the Nambu-Gorkov spinor as $\Phi$, we can identify the 
second spinor on the right hand side of (\ref{D34}) with the charge conjugate quasi-particle spinor, $\Phi_c= U_c \Phi$, so that this last equation
becomes
\begin{equation}
U(1)_{V,I} \otimes U(1)_{V,II}: \quad \Phi \to  \frac{e^{i\beta} + e^{i\alpha}}{2} \Phi 
+ \frac{e^{i\beta}-e^{i\alpha}}{2}  \Phi_c \, .
\label{D35}
\end{equation}
In this form, the equation is begging for introducing quasi-particle spinors of definite $C$-parity $\pm 1$ as
\begin{equation}
\Phi^{(\pm)} = \frac{\Phi \pm \Phi_c}{\sqrt{2}} \, ,
\label{D36}
\end{equation}
for which the transformations finally simplify to
\begin{equation}
U(1)_{V,I} \otimes U(1)_{V,II}:  \quad \Phi^{(-)} \to e^{i\alpha}\Phi^{(-)}, \quad \Phi^{(+)} \to e^{i\beta} \Phi^{(+)}  \, . 
\label{D37}
\end{equation}
This yields a physical interpretation of the pair of independent GN models equivalent to the sdGN model, consistent with the previous hint in Eq.~(\ref{D28c}):
They correspond to the two decoupled quasi-particle sectors with even and odd $C$-parity. 
The interactions implied by the self-dual choice of the Lagrangian apparently lead to 
the fact that not only $C$-parity is conserved, but that quasi-particles with opposite $C$-parity do not talk to each other.

The conserved fermion numbers of (\ref{D33}) can now be interpreted as numbers of quasi-particles with even or odd $C$-parity, which are 
separately conserved. In terms of $\phi_a$-components, they assume the form
\begin{equation}
Q^{(\pm)}  =   \frac{1}{2} \int dx \left( |\phi_1\pm \phi_3|^2+ |\phi_2 \mp \phi_4|^2 \right)\, .
\label{D38}
\end{equation}  
The total number of quasi-particles agrees with the norm of $\Phi$,
\begin{equation}
Q^{(+)} + Q^{(-)} = \int dx \left( |\phi_1|^2+ |\phi_2|^2 + |\phi_3|^2 + |\phi_4|^2 \right)\, .
\label{D39}
\end{equation}
Its conservation already follows from the hermiticity of $h$. 

We have only discussed the Noether charges. It is
easy to write down the corresponding Noether currents, starting from the vector currents of the GN models, 
\begin{eqnarray}
\rho^{(\pm)} & = & \frac{1}{2} \left( |\phi_1 \pm \phi_3|^2 + |\phi_2 \mp \phi_4 |^2 \right) \, ,
\nonumber \\
 j^{(\pm)} & = & \frac{1}{2} \left(- |\phi_1 \pm \phi_3|^2 + |\phi_2 \mp \phi_4 |^2 \right)\, ,
\nonumber \\
0 & = & \partial_0 \rho^{(\pm)} + \partial_1 j^{(\pm)} \, .
\label{D39a}
\end{eqnarray}

Coming back to the question of fermion number and solitons, we can now give the following answer: If we construct
a soliton solution of the sdGN model out of soliton solutions of the GN model with $N_I$ and $N_{II}$ fermions respectively,
it will carry $N^{(-)}=N_I$ ($N^{(+)}=N_{II}$) quasi-particles with odd (even) $C$-parity, respectively. The number of quasi-particles should not be confused with
fermion number, which is not a good quantum number in the sdGN model.

\section{Self-dual NJL model --- the perfect GN model}
\label{sect5}

In the last section, we have self-dualized the GN model by adding an interaction term where $\psi_1^{(i)}$ has been replaced by $\psi_1^{(i)*}$, 
the duality transformation. The resulting model then exhibits the discrete part of the Pauli-G\"ursey transformations (\ref{B5}) from the free,
massless theory. However, it breaks both U(1)$_V$ and U(1)$_A$ from the continuous chiral group explicitly. Here we apply the same procedure to the NJL model (\ref{C3})
by adding the dual interaction term from (\ref{C4}) to the original Lagrangian (\ref{C3}). The resulting model is unique in that the interaction
preserves the full Pauli-G\"ursey symmetry of the free Lagrangian. In this sense, it has maximal kinematic symmetry and has therefore been referred to as ``perfect" GN model
in Ref.~\cite{L37},
\begin{equation}
{\cal L}_{\rm pGN} = - 2i \psi_1^{(i)*} \partial \psi_1^{(i)} + 2 i \psi_2^{(i)*} \bar{\partial} \psi_2^{(i)}
+ 2 g^2 \left[ \left( \psi_1^{(i)*}\psi_2^{(i)} \right)  \left( \psi_2^{(j)*}\psi_1^{(j)} \right)
+ \left( \psi_1^{(i)*}\psi_2^{(i)*} \right)  \left( \psi_2^{(j)}\psi_1^{(j)} \right)\right]\, .
\label{E1}
\end{equation}
A similar model, but with two independent coupling constants in front of the fermion-antifermion and
fermion-fermion pairing term, is the subject of the recent Refs.~\cite{L11,L12} where also the self-dual special case (\ref{E1}) is touched upon. In these previous works
the emphasis was on the phase diagram of the models as a function of temperature and (vector and axial vector) chemical potentials $\mu, \mu_5$. 
Here we focus on the question of integrability and prepare the ground for solving soliton problems in the future. In our opinion the model (\ref{E1}) with a single
coupling constant is the only one in this family having any chance of being integrable.

Let us begin with a hint from literature in favor of the integrability of the pGN model. The four-fermion interaction in (\ref{E1}) can be
re-written in various ways (see also Refs.~\cite{L11,L12}) among which
\begin{equation}
{\cal L}_{\rm int} = g^2 \left( \psi_1^{(i)*} \psi_1^{(j)} - \psi_1^{(j)*}\psi_1^{(i)} \right) \left( \psi_2^{(i)*} \psi_2^{(j)} - \psi_2^{(j)*}\psi_2^{(i)} \right)\, .
\label{E2}
\end{equation}
In a celebrated paper where the dressing method was developed, Zakharov and Mikhailov have succeeded in mapping classical four-fermion models to principal
chiral models (PCM) known to be integrable \cite{L38}.  
They found that the $N$-flavor GN model corresponds to the Sp(2$N$,R) PCM, the NJL model to the U($N$) PCM. They also proposed a third fermionic
field theory related to the O($N$) PCM (occasionally referred to as Zakharov-Mikhailov (ZM) model \cite{L39}), which has not played any role in particle physics so far,
to the best of our knowledge. The four-fermion interaction of this model is identical to (\ref{E2}), so that we can identify the pGN model with the
quantum version of the ZM model. Since the classical integrability of the other two models survives quantization, it is 
plausible that this will also be true for the third model, although we cannot prove it. In contrast to the GN and NJL models, the classical Euler-Lagrange
equations of the ZM model do not have the same form as the TDHF equations of the quantum field theories, so that one cannot apply the dressing method
to find the solitons of the pGN model. In any case, the high degree of symmetry of the pGN model and the fact that the physics
is even richer than in the GN and NJL models due to the possibility of fermion-fermion pairing makes it seem worthwhile to try to 
actually solve the pGN model as well, irrespective of whether it is integrable or not.

In order to set up the HFB formulation of the pGN model, we can follow almost literally the steps performed in the previous section. The Hubbard-Stratonovitch
transformed Lagrangian now reads
\begin{eqnarray}
{\cal L}_{\rm pGN}' =  {\cal L}_{\rm pGN} - \frac{1}{2g^2} \left| \Delta + 2 g^2 \psi_1^{(i)*} \psi_2^{(i)} \right|^2
- \frac{1}{2g^2} \left| {\cal C} + 2 g^2 \psi_1^{(i)*} \psi_2^{(i)*} \right|^2 \, . 
\label{E3}
\end{eqnarray}
The auxiliary fields are complex and have been denoted as ($\Delta, {\cal C}$), to distinguish them from the real fields (${\cal S}, {\cal B}$) in 
the sdGN case. Expanding ${\cal L}_{\rm pGN}'$, we find
\begin{eqnarray}
{\cal L}'_{\rm pGN} & = & - 2i \psi_1^{(i)*}\partial \psi_1^{(i)} + 2 i \psi_2^{(i)*}\bar{\partial} \psi_2^{(i)}  - \Delta^* \psi_1^{(i)*}\psi_2^{(i)} -    \Delta \psi_2^{(i)*}\psi_1^{(i)}  
\nonumber \\
& & - {\cal C}^*  \psi_1^{(i)*}\psi_2^{(i)*}  -{\cal C} \psi_2^{(i)}\psi_1^{(i)}  - \frac{1}{2g^2} \left( |\Delta|^2 + |{\cal C}|^2 \right) \, ,
\label{E4}
\end{eqnarray}
with the constraint equations
\begin{eqnarray}
\Delta & = & - 2 g^2 \psi_1^{(i)*} \psi_2^{(i)} \, ,
\nonumber \\
{\cal C} & = & - 2 g^2  \psi_1^{(i)*} \psi_2^{(i)*}\, . 
\label{E5}
\end{eqnarray}
Evaluating the Hamiltonian density and writing the quantized Hamiltonian in the Nambu-Gorkov form (\ref{D6}), we obtain an expression 
analogous to (\ref{D6}), but with the 4$\times$4 matrix $h$ replaced by 
\begin{equation}
h = \left( \begin{array}{cccc} i \partial_1 & \Delta^* & 0 & {\cal C}^* \\  \Delta & -i \partial_1 & -{\cal C}^* & 0 \\
0 & -{\cal C} & i \partial_1 & -\Delta \\ {\cal C} & 0 & - \Delta^* & -i \partial_1 \end{array} \right) \, .
\label{E6}
\end{equation}
This is the first-quantized HFB Hamiltonian of the pGN model. For real condensates $\Delta= \Delta^* = {\cal S}, {\cal C}={\cal C}^* = {\cal B}$, 
it reduces to the Hamiltonian of the sdGN model. The TDHFB equations, $(h-i\partial_0)\Phi=0$, in (\ref{D26}) are replaced by
\begin{equation}
\left( \begin{array}{cccc} 2 i \partial & \Delta^* & 0 & {\cal C}^* \\ \Delta & - 2 i \bar{\partial} & -{\cal C}^* & 0 \\ 0 & -{\cal C} & 2i\partial & - \Delta \\
{\cal C} & 0 & - \Delta^*  & - 2i \bar{\partial} \end{array} \right) \left( \begin{array}{c} \phi_{1} \\ \phi_{2} \\
\phi_{3} \\ \phi_{4} \end{array} \right) = 0 \, , 
\label{E7}
\end{equation}
whereas the self-consistency conditions (\ref{D27}) go over into
\begin{eqnarray}
\Delta & = & -  Ng^2 \sum^{\rm occ} \left( \phi_1^* \phi_2  - \phi_4^* \phi_3 \right)\, ,
\nonumber \\
{\cal C} & = &  - Ng^2 \sum^{\rm occ} \left( \phi_1^* \phi_4  - \phi_2^* \phi_3  \right)\, .
\label{E8}
\end{eqnarray}
Due to the hermiticity of $h$, the norm (\ref{D39}) of $\Phi$, i.e., the total number of quasi-particles, is again conserved. This is the 
Noether charge of the symmetry transformation $\phi_i \to e^{i\alpha} \phi_i, i=1,...,4$ with the conserved current
\begin{eqnarray}
\rho & = & |\phi_1|^2 + |\phi_2|^2 + |\phi_3|^2 + |\phi_4|^2 \, ,
\nonumber \\
j & = & -|\phi_1|^2 + |\phi_2|^2 - |\phi_3|^2 + |\phi_4|^2 \, .
\label{E8a}
\end{eqnarray}
However, $h$ is no longer invariant under charge
conjugation, but we have
\begin{equation}
U_c h(\Delta, {\cal C}) U_c^{\dagger} = h(\Delta^*, {\cal C}^*)
\label{E9}
\end{equation}
instead.
As a consequence the second continuous symmetry which we had found in the sdGN case is not present here. We also cannot
block-diagonalize $h$ by a constant, unitary transformation, so that the solutions of the pGN model can in general
not be reduced to those of any simpler, integrable model. This increased complexity can also be seen if we express the 
Lagrangian of the pGN model in terms of Majorana spinors, as in (\ref{D19}). The result is 
\begin{equation}
{\cal L}_{\rm pGN} =  {\cal L}_{\rm sdGN} - g^2  \left(\chi_1^{(i)} \chi_4^{(i)}\right)^2 - g^2 \left(\chi_3^{(i)} \chi_2^{(i)}\right)^2\, .
\label{E10}
\end{equation}
The extra terms introduce interactions between the two independent GN models in (\ref{D19}) and destroy the trivial solubility of the model.
Of course, this does not rule out that the perfect GN model is also integrable, but one needs to work harder.

We have not yet been able to solve the TDHFB equations for soliton solutions in any systematic way, and leave this to the future.
Nevertheless, it is possible to give a few examples of self-consistent solutions. We first note that any solution of the sdGN model
is also a solution of the pGN model. This is a non-trivial statement because of the different self-consistency conditions in both
cases. Thus for instance, a solution of the ordinary GN model does not solve the NJL model in general, unless the total fermion number
vanishes \cite{L23}. The way how it works in the case of the two self-dual models is as follows. We start from a solution of the
sdGN model with spinors in the form given in Eqs.~(\ref{D28a},\ref{D28b}). If we plug those spinors into the self-consistency conditions
(\ref{E8}) and sum over occupied states, we find that $\Delta=(S_{II}+S_I)/2={\cal S}, {\cal C} = (S_{II}-S_I)/2={\cal B}$, i.e., real condensates.  

Starting from these solutions with real mean fields, we can then try to make the potentials complex by a unitary transformation,
changing the phases of the spinor components. If these phases depend on $z,\bar{z}$, they will destroy the form of the TDHFB equations
because the derivatives act on the phases. The only exceptions are complex phases which are  are either constant or linearly $x$ dependent. These special cases
give us a first handle on the soliton problem and hopefully will be useful for eventually constructing a general solution and proving (or disproving)
integrability of the pGN model. Let us consider these two possibilities.
\vskip 0.2cm
a) {\em Constant phases}
\vskip 0.2cm
Suppose that the mean fields have the form
\begin{equation}
\Delta = e^{i\alpha} {\cal S}, \quad {\cal C} = e^{i\beta} {\cal B}\, ,
\label{E11}
\end{equation}
where $\alpha, \beta$ are constant but ${\cal S}, {\cal B}$ real functions of $x,t$. In this case, the complex phases can be eliminated
by the constant unitary transformation
\begin{equation}
U = {\rm diag} \left( e^{i(\alpha+\beta)/2},  e^{-i(\alpha-\beta)/2},  e^{-i(\alpha+\beta)/2},  e^{i(\alpha-\beta)/2} \right)   
\label{E12}
\end{equation} 
in the form
\begin{equation}
U h(\Delta,{\cal C}) U^{\dagger} = h({\cal S},{\cal B})\, .
\label{E13}
\end{equation}
There are two physical examples where this method can be applied: The vacuum and the twisted kink between superconducting and chirally broken
phases.

Take the vacuum first. If the condensates are constant, they can be parameterized as 
\begin{equation}
\Delta = e^{i\alpha} m, \quad {\cal C} = e^{i\beta} M\, ,
\label{E14}
\end{equation} 
with real, non-negative ($m,M$). The unitary transformation (\ref{E12}) maps the vacuum problem of the pGN model onto that of the sdGN model.
The four discrete vacua of the sdGN model of Fig.~\ref{Fig1} go over into the vacuum manifold of the pGN model consisting of two disconnected unit circles, in natural units
(the chiral circle and the circle of the Cooper pair condensate).
Which point is chosen on which circle is physically irrelevant, as always in SSB.
The renormalized vacuum energy density of the pGN model is the same as that of the sdGN model.

A more interesting result of this unitary transformation arises if we apply it to the domain wall between chirally broken and superconducting 
phases of the sdGN model, i.e., the kink of Fig.~\ref{Fig2} and Eq.~(\ref{D29}). Under the inverse of the unitary transformation (\ref{E12}), this goes over into
\begin{equation}
\Delta = \frac{1}{2} (1+\tanh x )e^{i\alpha}, \quad {\cal C} = \frac{1}{2} (1-\tanh x)e^{i\beta}\, ,
\label{E15}
\end{equation}
interpolating between the superconducting vacuum ($\Delta=0, {\cal C}=e^{i\beta}$) at $x \to - \infty$
and the chirally broken vacuum ($\Delta=e^{i\alpha},{\cal C}=0$) at $x\to \infty$. 
This is a new kind of twisted kink and an exact self-consistent soliton solution of the HFB equations for the pGN model. Notice that the width of the kink 
does not depend on the twist angles, unlike the twisted kink of the original NJL model \cite{L40}.

If we apply the same transformation to the more complicated domain wall of Figs.~\ref{Fig3}, \ref{Fig4} and Eq.~(\ref{D30}), we arrive at the condensates
\begin{equation}
\Delta   =  \frac{1}{2} \left[ \tanh (x-a) +\tanh (x+a) \right]e^{i\alpha} , \quad {\cal C} = \frac{1}{2} \left[ \tanh (x-a) - \tanh (x+a) \right]e^{i\beta} \, .
\label{E16}
\end{equation}
The phase $\beta$ allows us to dial the phase of the Cooper pair condensate inside the kink region. However, the vacua at $x\pm \infty$ are always
located at diametrically opposing points on the chiral circle, $\pm e^{i\alpha}$. It is not possible to generate in such a manner the most general kink, which should depend
on three different twist angles. 

\vskip 0.2cm
b) {\em Linearly $x$-dependent phases}
\vskip 0.2cm
In the NJL model, a chiral transformation with a linearly $x$-dependent phase generates at the same time a chemical potential
(from the spatial derivatives) and a helical condensate (``chiral spiral" \cite{L41}). As is well known, this leads to a crystalline structure of
cold and dense matter. We can copy this trick here. Depending on whether the vacuum has chiral symmetry breaking or 
Cooper pairs, we will be dealing with the same phenomenon as in the NJL model, or with an inhomogeneous 
superconductor (the Larkin-Ovchinnikov-Fulde-Ferrel (LOFF) phase \cite{L42,L43}). Consider first the case where the vacuum has $m=1,M=0$ and define the unitary transformation
\begin{equation}
U_1 = {\rm diag} \left( e^{-i\mu x}, e^{i\mu x}, e^{i\mu x}, e^{-i\mu x} \right) \, .
\label{E17}
\end{equation} 
The unitary transformation of $h$ yields
\begin{equation}
U_1 \left( \begin{array}{cccc} i \partial_1 & 1 & 0 & 0 \\ 1 & -i\partial_1 & 0 & 0 \\ 0 & 0 & i\partial_1 & -1 \\ 0 & 0 & -1 & -i \partial_1 \end{array} \right) U_1^{\dagger}
= \left( \begin{array}{cccc} i\partial_1-\mu & e^{-2i\mu x} & 0 & 0 \\ e^{2i\mu x} & -i \partial_1 -\mu & 0 & 0 \\ ß & 0 & i\partial_1 + \mu & -e^{2i\mu x} \\ 0 & 0 & -e^{-2i\mu x} & -i\partial_1 + \mu 
\end{array} \right)\, .
\label{E18}
\end{equation}
This corresponds to introducing a vector chemical potential $\mu$ and condensates in the form of the standard chiral spiral, 
$\Delta=e^{2i\mu x}$. If the vacuum is superconducting ($m=0,M=1$),
we choose the unitary transformation
\begin{equation}
U_2 = {\rm diag} \left( e^{i\mu_5 x}, e^{i\mu_5 x}, e^{-i\mu_5 x}, e^{-i\mu_5 x} \right) \, .
\label{E19}
\end{equation} 
We then map the Cooper pair vacuum onto the LOFF state with ${\cal C} = e^{-2i\mu_5 x}$,
\begin{equation}
U_2 \left( \begin{array}{cccc} i \partial_1 & 0 & 0 & 1 \\ 0 & -i\partial_1 & -1 & 0 \\ 0 & -1 & i\partial_1 & 0 \\ 1 & 0 & 0 & -i \partial_1 \end{array} \right) U_2^{\dagger}
= \left( \begin{array}{cccc} i\partial_1+\mu_5  & 0 & 0 & e^{2i\mu_5 x} \\ 0 & -i \partial_1 - \mu_5 & -e^{2i\mu_5 x} & 0 \\ 0 & -e^{-2i\mu_5 x} & i\partial_1 - \mu_5 & 0 \\ e^{-2i\mu_5 x} & 0 & 0
 & -i\partial_1 + \mu_5 \end{array} \right) \, .
\label{E20}
\end{equation}
Now $\mu_5$ has to be interpreted as axial chemical potential, since it enters with opposite sign for left- and right-handed fermions.
These spiral states have also been discussed in \cite{L12} in the more general model with two coupling constants
and at finite temperature. Since only one type of condensates appears, these structures are identical to what one expects in the  NJL model and its dual
discussed in Sec.~\ref{sect3}. 

Besides these cases, there are a number of trivial solutions of the pGN model. If we choose for ($\phi_1, \phi_2$) any solution of the NJL model
with self-consistent potential $\Delta$ and set $\phi_3 = \phi_4 = 0$, this yields a self-consistent solution of the pGN model as well. The
system then does not take advantage of the possibility of fermion-fermion pairing at all. Likewise, we can introduce any solution of the NJL model
with potential $\Delta$ into the ($\phi_1, \phi_4$) components and set $\phi_2=\phi_3=0$. This yields a solution of the pGN model with Cooper pair
condensate ${\cal C} = \Delta$, but vanishing chiral condensate.  However, we have not yet found
any soliton solution where both $\Delta$ and ${\cal C}$ are non-vanishing, other than the twisted kinks above. If the pGN model is indeed integrable, we
would expect that such solutions should exist in closed analytical form, by analogy with the other integrable models.

\section{Summary and outlook}
\label{sect6}

For many years, integrability of the massless GN and chiral GN models seemed like a rather academic issue. The derivation of hadron masses
(mesons, baryons, multi-baryon bound states) and of the phase diagrams in the ($T,\mu$) plane could equally well be done for the massive case 
\cite{L44,L45,L46,L47,L48,L49,L50} as for the massless case (\cite{L19,L33,L34,L40,L42,L51}, even to a large extent analytically, although only the massless
models are integrable.
With the study of time dependent problems, this perspective has changed in recent years. Scattering of baryons for instance could only be solved
in the massless GN and NJL models. Properties characteristic for integrable systems show up most clearly in scattering processes ---
transparent self-consistent potentials, factorization of transmission amplitudes, additivity of masses of bound states \cite{L20,L21,L22}. 

These findings have incited us to think about other potentially integrable four-fermion models. From the strong interaction physics point of view, models
giving rise to Cooper pairing are particularly interesting as toy models for color superconductivity in QCD. In this paper we have indeed identified
three such models which are definitely integrable and one model which is a candidate, but for which there is no proof yet.  Two 
out of these models are rather trivial: By replacing the fermion-antifermion interaction in the standard GN models by fermion-fermion interactions
using a simple duality transformation (particle-hole conjugation of fermions with one chirality only), the standard GN and NJL models are mapped onto
two Cooper pair models with real and complex Cooper pair condensates, respectively. The second one, the dual NJL model, has already been studied in the 
literature some time ago \cite{L7}. Both models can be trivially solved and shown to be integrable by noting that the duality transformation is canonical, so that it is
only a matter of interpretation whether one talks about chiral symmetry breaking or superconductivity. This reminds us of the title of the original NJL 
paper, {\em Dynamical  model of elementary particles based on an analogy with superconductivity} \cite{L2}. 

More interesting candidates for integrable models have been obtained by self-dualizing
the starting models, i.e., adding the dual (fermion-fermion) to the original (fermion-antifermion) pairing interaction. The motivation behind these attempts is that 
we expect integrable models to have only one coupling constant and particularly high symmetry. This does not leave much choice. If we 
self-dualize the GN model, surprisingly we arrive at a model equivalent to a pair of independent GN models. While we obtained this result using 
the HFB approach, with hindsight one can see this decoupling already at the Lagrangian level, provided one formulates it with Majorana fields rather
than Dirac fields. Physically, the fermions of the two independent GN models are closely related to quasi-particles with definite $C$-parity of the 
sdGN model.

Perhaps the most intriguing candidate for an integrable model is the self-dual NJL model. It is unique in the sense that it shares the full Pauli-G\"ursey
symmetry with the free, massless Dirac theory, i.e., has maximal kinematic symmetry. This is why it was referred to as perfect GN model \cite{L37}.
Classically this model reduces to a model proposed by Zakharov and Mikhailov \cite{L38} which can be mapped onto the PCM of the orthogonal group.
We take this
as a hint of integrability of the quantum theory as well. However, so far we could find only soliton solutions which can be reduced to known ones from
the GN models or the sdGN model. Genuine solitons with both fermion-fermion and fermion-antifermion condensates still have to be found. 
In view of the increased complexity of the TDHFB as compared to the TDHF approach, this is actually quite a challenge. It remains
to be seen whether the methods developed for solving the GN and NJL models in Refs.~\cite{L19,L20,L21,L22} can be generalized to this situation. 
This would enable us to confirm or disprove integrability of the pGN model.


\end{document}